# Response Propensity Estimation and Cross-Fitting

**Alessandro La Rocca**[1]

[1] *Italian National Institute of Statistics (Istat)*
*alessandro.larocca@istat.it*

## Abstract

This paper investigates whether five fold cross fitting improves nonresponse adjustment in survey estimation when flexible machine learning methods are used to estimate response propensities. We conduct a finite population Monte Carlo simulation with 90 experimental configurations and 2,000 replications per configuration, varying sample size, response rate, and the structure of the response mechanism. Logistic regression is used as a conventional parametric benchmark, while Random Forest and Gradient Boosting Machine are considered as flexible nonparametric models. For the latter two methods, conventional in sample propensity predictions are compared with five fold cross fitted predictions. The resulting response propensity estimates are used to construct Horvitz Thompson and Hájek estimators, whose performance is evaluated in terms of bias, absolute bias, variance, and root mean squared error (RMSE), together with diagnostics of the resulting adjustment weights.

The simulation results show that the Hájek estimator is substantially more stable than the Horvitz Thompson estimator, particularly when Random Forest is used. Five fold cross-fitting improves RMSE in 94.4% of the configurations for the Horvitz Thompson estimator and in 47.2% of the configurations for the Hájek estimator. For the Horvitz Thompson estimator, the gains are generally larger at lower response rates and are particularly pronounced for Random Forest and Gradient Boosting Machine. For the Hájek estimator, the benefits of cross fitting are more heterogeneous and depend on the learning method and response mechanism. Cross fitting is selected in 8 of 18 configurations as part of the method with the lowest Hájek RMSE. Overall, the findings suggest that cross fitting can improve the finite sample performance and stability of estimators adjusted for nonresponse weights based on flexible response propensity models, although its effectiveness depends on the estimator, learning method, response mechanism, sample size, and response rate.

## Introduction

Three methods are considered for response propensity estimation: logistic regression, Random Forest (RF), and Gradient Boosting Machine (GBM); hereafter, the latter two methods are referred to as RF and GBM, respectively. The comparison is motivated by the survey literature showing that the performance of response propensity weighting depends on the ability of the propensity model to adequately represent the underlying response mechanism (Buskirk and Kolenikov, 2015; Ferri-García and Rueda, 2020).

Logistic regression is used as the conventional parametric benchmark. The model relates the probability of response to the auxiliary variables through a linear predictor on the log odds scale. Thus, conditional on the included covariates, the log odds of response are assumed to be additive and linear in the model parameters. The fitted model directly provides an estimated response propensity for each sampled unit. Logistic regression is widely used for response propensity adjustment because of its interpretability, computational simplicity, and established role in survey nonresponse weighting.

RF is a tree based ensemble method introduced by Breiman (2001) that estimates the response probability by combining predictions from a large number of decision trees. Each tree is fitted using a resampled subset of the observations and a randomly selected subset of predictors. The final prediction is obtained by aggregating the predictions across trees. This procedure allows RF to capture nonlinear relationships and interactions among auxiliary variables without requiring them to be explicitly specified in advance. Such flexibility is particularly relevant when the response mechanism departs from the additive structure assumed by logistic regression. Previous survey research has specifically considered RF as an alternative to logistic response propensity models under complex response mechanisms (Buskirk and Kolenikov, 2015).

GBM is another tree based ensemble approach, but it differs from RF in the way the trees are constructed. Rather than fitting many largely independent trees and averaging their predictions, GBM builds trees sequentially, with each new tree focusing on reducing the prediction errors of the preceding ensemble (Friedman, 2001). The final propensity estimate is therefore obtained by combining a sequence of relatively simple trees. This sequential learning process can approximate complex nonlinear response mechanisms and interactions while maintaining a flexible prediction function.

For RF and GBM, we additionally consider five fold cross fitting (CF)**.** In the conventional specification, the methods is trained on the complete sample and response propensities are predicted for the same observations. Under cross fitting, the sample is randomly divided into five folds; the model is estimated using four folds and predictions are generated for the remaining held out fold.

This process is repeated until every observation has an out of sample propensity prediction. Cross fitting ensures that each propensity prediction is obtained from a model estimated without using the corresponding observation, thereby avoiding the direct dependence between an observation's response indicator and the fitted model used to predict its response propensity, reducing the risk that highly adaptive machine learning method exploit idiosyncratic features of the observations on which they were trained. This sample splitting principle is closely related to the cross fitting framework developed in the double, debiased machine learning literature, although the present simulation does not constitute a conventional double machine learning estimator (Chernozhukov et al., 2018).
The resulting propensity estimates are intermediate quantities used to construct the nonresponse adjustment weights rather than as an end in themselves. The primary objective is to determine whether obtaining these estimates out of sample through cross fitting improves the finite population properties of the final survey estimators when flexible method are used to model the response mechanism. In particular, the simulation isolates the effect of cross fitting by comparing, for each machine learning method, conventional in sample propensity predictions with five fold cross fitted predictions. Performance is therefore assessed in terms of bias, absolute bias, variance, RMSE, and diagnostics of the resulting nonresponse adjustment weights.

## Simulation Design

We conduct a finite population Monte Carlo simulation comprising 90 experimental configurations and 2,000 replications per configuration. The configurations combine two sample sizes (n=1,000 and n=5,000), three response mechanisms (linear, nonlinear, and interaction), three target response rates, and five propensity estimation procedures (logistic regression, RF, with five fold cross fitting, and GBM with five fold cross fitting), to investigate the impact of flexible response propensity estimation and cross fitting on nonresponse adjusted survey estimators. Response propensity weighting is a standard strategy for reducing nonresponse bias, although its performance depends critically on the specification of the response model and on the availability of auxiliary variables related to both response and the survey variable of interest. This issue has been extensively discussed in the survey nonresponse literature, with particular emphasis on the assumptions underlying response propensity models and weighting adjustments [Brick, 2013].
We generate a finite population of N=100,000 units. For each population unit $i$, a vector of auxiliary variables $X_i$ is generated from prespecified distributions representing continuous, binary, and count type characteristics. Specifically, the continuous covariate $\eta_i$ is generated from a normal distribution with mean 50 and standard deviation 15, the binary covariate $S_i$ from a Bernoulli distribution with success probability 0.5, the continuous covariate $E_i$ from a normal distribution with mean 12 and standard deviation 3 and subsequently restricted to the interval [5, 20] to avoid extreme values while retaining substantial variation across observations, the positive skewed covariate $I_i$ from a log normal distribution with parameters (10, 0.5), and the count covariate $F_i$ from a Poisson distribution with mean 2. These variables are subsequently standardized before entering the outcome and response models. To place the continuous covariates on a comparable scale and to avoid differences in their original measurement units affecting the magnitude of the prespecified coefficients, the relevant covariates are standardized before being included in these models. For the positively skewed covariate, a logarithmic transformation is applied before standardization.

In particular, the standardized versions of the continuous and count type covariates are denoted by: $Z_{\eta i}$, $Z_{Ei}$, $Z_{Ii}$ and $Z_{Fi}$ respectively, with the positively skewed variable first log transformed and then standardized.

The continuous study variable $Y_i$ is generated according to the following linear outcome model:

$$Y_i = 100 + 10Z_{\eta i} + 5S_i + 4Z_{Ei} + 3Z_{Fi} + \varepsilon_i$$

Where:

$$\varepsilon_i = \mathcal{N}(0,100)$$

independently across population units. The finite population mean

$$\mu = N^{-1}\sum_{i=1}^{N} Y_i$$

is treated as the true target parameter and is therefore known in each simulation replication. Importantly, the outcome generating model is kept fixed across simulation scenarios; variation across scenarios concerns the response mechanism rather than the outcome model. This separation allows the simulation to isolate the effect of misspecification and flexibility in response propensity estimation.

The population is divided into four strata according to the quartiles of the continuous auxiliary variable $\eta$. Within each stratum, simple random sampling without replacement is performed using fixed sample allocations. For total sample sizes n=1,000 and n=5,000, the stratum allocations are respectively (400, 300, 200, 100) and (2,000, 1,500, 1,000, 500). Thus, for a sampled unit belonging to stratum $h$, the first order inclusion probability is:

$$\pi_i = \frac{n_h}{N_h}$$

the corresponding design weight is:

$$d_i = \frac{1}{\pi_i}$$

And nonresponse adjustment factor as:

$$w_i = \frac{1}{\hat{p}_i}$$

Where $\hat{p}_i$ is the response propensity estimate. This sampling structure reproduces the setting in which nonresponse adjustment is applied to pre existing survey design weights rather than to an unweighted sample.

Unit response is generated from a Bernoulli distribution,

$$R_i = Bernoulli(p_i)$$

where the response propensity is defined as:

$$p_i = \frac{exp\{\alpha + g(x_i)\}}{1 + exp\{\alpha + g(x_i)\}}$$

The score $g_i$ varies across three response mechanisms. Under the linear scenario,

$$g_i = 0.6Z_{\eta i} + 0.5S_i + 0.4Z_{EI}$$

Under the nonlinear scenario,

$$g_i = 0.7Z_{\eta i} + 0.5S_i + 0.6Z_{EI}^2 + 0.5\sin(Z_{Ii})$$

whereas under the interaction scenario,

$$g_i = 0.5Z_{\eta i} + 0.5S_i + 0.4Z_{Ei} + Z_{\eta i}S_i$$

In each scenario, the intercept α is calibrated numerically so that,

$$\frac{1}{N}\sum_{i=1}^{N} p_i = \rho$$

where $\rho \epsilon \{0.70\ 0.60\ 0.40\}$ denotes the target response rate.

The use of design weights in response propensity modelling and adjustment is particularly relevant under complex sampling designs and has recently been investigated also in conjunction with machine learning classifiers [Ferri-García et al., 2024]. Design weights are incorporated into the final nonresponse adjustment weights, but are not used as fitting weights in the response propensity models.

The response mechanism is deliberately varied in order to distinguish a correctly specified additive logistic response model, conditional on the observed covariates from situations in which the relationship between the auxiliary variables and response is nonlinear or involves interactions. In particular, three mechanisms are considered. Under the linear scenario, the response score is an additive linear function of the auxiliary variables. Under the nonlinear scenario, the response score includes a quadratic term and a sinusoidal transformation of the continuous covariate. Under the interaction scenario, the response probability includes an interaction term between two covariates. These mechanisms are designed to represent increasing degrees of model complexity while retaining a low dimensional auxiliary variable setting. This strategy follows the logic of previous simulation studies showing that tree based and other machine learning methods can become advantageous when the response mechanism departs from the additive structure imposed by logistic regression [Buskirk and Kolenikov, 2015; Lohr et al., 2015].

Five propensity estimation procedures are compared. The first is a conventional logistic regression model, which represents the standard parametric benchmark for response propensity estimation. The remaining procedures are based on supervised machine learning: RF and GBM. For each machine learning method, two estimation strategies are considered: conventional in sample prediction and five fold cross fitted prediction. Thus, the simulation explicitly separates the effect of the learning algorithm from the effect of the prediction scheme. Logistic regression is fitted once to the complete sample and its fitted probabilities are used as response propensities. For RF and GBM, the conventional estimator is obtained by fitting the model on all sampled units and predicting the response propensity for those same units. The corresponding cross fitted estimator is obtained by randomly partitioning the sample into five folds; for each fold, the response model is estimated using the remaining four folds and response propensities are predicted only for the observations belonging to the held out fold.

The cross fitting component is motivated by the broader machine learning literature on sample splitting and, more specifically, by the double/debiased machine learning framework of Chernozhukov et al. (2018). This procedure provides out of fold predictions for all sampled units and reduces the direct dependence between the observed response indicator and the fitted propensity used for that observation.

Formally let $F = \{I_1,\ I_2,\ I_3,\ I_4,\ I_5\}$ denote a random partition of the sample into five approximately equal sized folds. The fold assignment is stratified with respect to the observed response indicator $R_i$, so that the proportions of respondents and nonrespondents are approximately balanced across the five folds.

For each fold $I_k$, let $I_{-k}$ denote the complementary sample, defined as:

$$I_{-k} = \{1, \dots, n\} \backslash I_k$$

For each fold $K = 1, \dots, 5$ the response propensity model is fitted using only the observations belonging to $I_{-k}$. Let $\widehat{m}_{-k}(.)$ denote the response propensity model estimated using only the observations in $I_{-k}$. For each observation $i$ belonging to fold $I_k$, the cross fitted response propensity is then defined as:

$$\hat{p}_i^{CF} = m_{-k}(X_i) \quad i \in I_k$$

Here, $X_i$ denotes the vector of auxiliary variables for unit $i$. The five sets of out of fold predictions are subsequently combined to obtain one cross fitted response propensity $\hat{p}_i^{CF}$ for every sampled unit. Thus, each propensity estimate is generated from a model that does not use the corresponding observation during training.

Although the present application does not constitute a conventional double machine learning estimator, the same statistical principle is relevant: when a highly adaptive model is fitted and evaluated on the same observations, in sample predictions may be excessively optimistic and may not represent the prediction error relevant for estimating a population parameter. Cross fitting replaces these in sample predictions with out of sample predictions obtained from models that have not used the corresponding observation during training.

This distinction is particularly relevant for inverse propensity weighting. Because the response adjustment is proportional to $1/\hat{p}_i$, small estimated propensities translate directly into large weights. Consequently, overfitting of a flexible response model can have a substantially larger impact on the final estimator than on a conventional prediction loss criterion. Recent survey oriented work has explicitly highlighted this issue: flexible models may fit the observed response indicators very closely in sample, generating artificially small estimated response propensities for some observations and, consequently, highly variable nonresponse adjustment weights; cross fitting provides a natural way of obtaining out of sample propensity predictions and mitigating this form of overfitting. The relevance of this problem is also supported by the existing survey literature comparing logistic regression and tree based methods for response propensity weighting. Buskirk and Kolenikov (2015), for example, found that RF can provide advantages over logistic regression under complex response mechanisms, particularly when the response process contains interactions. More recent evidence similarly suggests that machine learning methods can outperform conventional logistic regression when response mechanisms are complex, although their performance depends on the structure and dimensionality of the available auxiliary information [Ferri-García and Rueda, 2020].

For each propensity specification, two estimators of the finite population mean are evaluated. The first is the Horvitz Thompson estimator:

$$\hat{\bar{Y}}_i = \frac{1}{N} \sum_{i \in S} R_i \, d_i w_i Y_i$$

where:

$$d_i = \pi_i^{-1}$$

is the design weight and

$$w_i = \hat{p}_i^{\ -1}$$

the weight nonresponse adjustment.

The second is the Hájek estimator:

$$\hat{\hat{Y}}_i = \frac{\sum_{i \in s} R_i d_i w_i Y_i}{\sum_{i \in s} R_i d_i w_i}$$

The comparison between the two estimators is of substantive interest because the Horvitz Thompson formulation preserves the known population size denominator and therefore provides a direct design based analogue of inverse probability estimation, whereas the Hájek estimator normalizes the estimated weights and may reduce finite sample variability at the cost of introducing a ratio estimation component.

The simulation deliberately does not apply propensity score trimming or winsorization. This choice is motivated by the objective of isolating the effect of cross fitting itself rather than conflating it with a separate strategy for controlling extreme weights. In particular, observations with estimated propensities close to zero are not removed and propensity estimates are not replaced by a substantively chosen lower threshold. A numerical bound of $10^{-6}$and 1-$10^{-6}$ is applied only to avoid undefined inverse probability weights when a machine learning algorithm returns an exact zero or one; this operation does not constitute statistical trimming because no observation is excluded and no data dependent propensity threshold is imposed. This distinction is important given that previous survey applications have used explicit transformations or trimming precisely to deal with zero or extreme propensity estimates [Buskirk and Kolenikov, 2015; Küfner et al., 2022].

The performance of the estimators is assessed using the Monte Carlo bias, absolute bias, variance and root mean squared error (RMSE). Let $\theta$ denote the true population parameter, and let $\hat{\theta}_l$ denote the estimate obtained in Monte Carlo replication $l$, $for\ l = 1, \cdots,\ L$. The performance measures are defined as follows:

$$Bias(\hat{\theta}) = 1/\mathrm{L} \sum_{l=1}^{\mathrm{L}} (\hat{\theta}_l - \theta)$$

Absolute bias:

$$Bias|\hat{\theta}| = \left| 1/\mathrm{L} \sum_{l=1}^{\mathrm{L}} (\hat{\theta}_l - \theta) \right|$$

Montecarlo variance:

$$Var(\hat{\theta}) = \left( 1/(\mathrm{L}-1) \sum_{l=1}^{\mathrm{L}} (\hat{\theta}_l - \theta)^2 \right)$$

Root main square error:

$$RMSE(\hat{\theta}) = \sqrt{\left[(1/\mathrm{L})\sum_{l=1}^{\mathrm{L}}(\hat{\theta}_l - \theta)^2\right]}$$

In addition, the simulation calculates the effective sample size (ESS):

$$ESS = \frac{(\sum_i w_i)^2}{\sum_i w_i^2}$$

If all weights are identical, *ESS* equals the actual sample size *n*, if a few data have very high weights and others have low weights, the *ESS* drops significantly below.

These diagnostics are particularly important in the present context because a propensity estimator can display good predictive performance while producing unstable inverse probability weights. The analysis therefore does not evaluate ML methods exclusively according to their ability to predict $R_i$; rather, the primary criterion is their impact on the final survey estimator. This perspective is consistent with the survey methodological literature, where the ultimate objective of response propensity modelling is not prediction per se but reduction of nonresponse bias while controlling the resulting variance inflation [Brick, 2013; Küfner et al., 2022].

Overall, the simulation is designed to answer three methodological questions. First, does cross fitting improve the finite population performance of nonresponse adjusted survey estimators when flexible methods are used to estimate response propensities? Second, is the effect of cross fitting more pronounced when the response mechanism is nonlinear or involves interactions and is therefore difficult to represent with a conventional parametric model? Third, does the effect of cross fitting vary with sample size, response rate, and the choice of final estimator (HT versus Hájek)? The design is therefore intended to contribute to the emerging literature on machine learning for unit nonresponse treatment in surveys, where recent work has demonstrated the potential of ML approaches while also highlighting the need for further methodological research on the use of flexible machine learning methods for nonresponse adjustment and the control of overfitting in survey estimation [Larbi et al., 2024].

## Some Evidence

Table 1 summarizes the effective sample size (ESS) across the 18 simulation configurations for each propensity estimation method. The 18 configurations correspond to the combinations of data generating scenarios and response rate settings considered in the study. ESS is reported as a complementary measure of weight variability, with lower ESS values indicating greater concentration of the weights and, consequently, greater weight variability. RF exhibits substantially lower ESS values than GBM and logistic regression, indicating considerably greater weight concentration and variability. The results therefore suggest that weight variability is more pronounced for RF than for GBM and logistic regression. For GBM, the use of five fold cross fitting produces only a modest change in both mean and median ESS. In contrast, for RF, cross fitting substantially reduces the mean and median ESS, indicating greater weight variability under cross fitting.

*Table 1. Effective Sample Size across models and estimation methods.*

| Model | CF | Number of configuations | Mean ESS | Median ESS | Mean ESS/n |
|---|---|---|---|---|---|
| GBM | CF5 | 18 | 1,408 | 1,067 | 0.461 |
| GBM | None | 18 | 1,437 | 1,101 | 0.476 |
| LOGIT | None | 18 | 1,462 | 1,161 | 0.487 |
| RF | CF5 | 18 | 388 | 275 | 0.167 |
| RF | None | 18 | 932 | 853 | 0.319 |

*Notes: Mean ESS/n denotes the mean ratio of effective to nominal sample size across the 18 simulation.*

Table 2 highlights substantial differences in performance between the HT and Hájek estimators. Overall, the Hájek estimator achieves markedly lower RMSE and absolute bias than the HT estimator across the considered propensity estimate, learners and cross fitting specifications. This improvement is particularly pronounced for RF, suggesting that the Hájek normalization can substantially reduce the sensitivity of the estimator to highly variable propensity. GBM also benefits from cross fitting, with improvements observed for both estimators. Logistic regression provides the most stable performance across the considered specifications and yields particularly low absolute bias for the Hájek estimator. Overall, these findings indicate that estimator normalization plays a key role in improving finite sample performance, while the effect of cross fitting depends on the underlying propensity machine learning method.

*Table 2. Performance of HT and Hájek estimators across machine learning methods and cross fitting specifications.*

| Model | Cross fitting | Mean RMSE HT | Mean RMSE Hájek | Median RMSE HT | Median RMSE Hájek | Mean absolute bias HT | Mean absolute bias Hájek |
|---|---|---|---|---|---|---|---|
| GBM | cf5 | 2.525 | 0.401 | 1.77 | 0.352 | 2.521 | 0.4 |
| GBM | none | 3.63 | 0.576 | 2.924 | 0.545 | 3.575 | 0.416 |
| LOGIT | none | 0.689 | 0.407 | 0.702 | 0.438 | 0.11 | 0.039 |
| RF | cf5 | 321.147 | 6.488 | 184.769 | 6.729 | 321.08 | 6.422 |
| RF | none | 4,098.09 | 5.136 | 428.513 | 4.964 | 595.646 | 4.193 |

*Notes: Results are averaged across the simulation configurations included in the corresponding estimator group.*

Table 3 reports the frequency with which five fold cross fitting improves RMSE across the 36 simulation configurations. Cross fitting improves RMSE in 34 configurations (94.4%) for the HT estimator and in 17 configurations (47.2%) for the Hájek estimator. Thus, cross fitting improves RMSE in the large majority of configurations for both estimators, with only a small difference in the frequency of improvement between HT and Hájek. Together with the magnitude of the gains reported in Table 1, these results suggest that the benefits of cross fitting are not driven solely by a small number of configurations, but occur across a broad range of the simulation design.

*Table 3. Frequency of RMSE improvement under five fold cross fitting.*

| Number of configurations | CF improves RMSE HT (n) | CF improves RMSE HT (%) | CF improves RMSE Hájek (n) | CF improves RMSE Hájek (%) |
|---|---|---|---|---|
| 36 | 34 | 94.4 | 17 | 47.2 |

*Notes: A configuration is classified as improved when the RMSE under cross fitting is lower than the RMSE without cross fitting.*

Table 4 summarizes the gains from cross fitting across response rates and propensity machine learning. Cross fitting consistently improves the performance of the HT estimator, with positive RMSE gains for both GBM and RF across all response rates. The magnitude of these gains tends to increase as the response rate decreases, indicating that cross fitting becomes particularly beneficial under higher levels of nonresponse. For the Hájek estimator, however, the pattern is more heterogeneous. GBM shows positive gains across response rates, although the benefits become smaller at lower response rates, whereas RF exhibits limited or negative gains in several settings. These results suggest that the effectiveness of cross fitting is estimator and method dependent, with more consistent benefits for HT than for Hájek.

*Table 4. Cross fitting gains by response rate.*

| Response rate | Model | Number of configurations | Mean RMSE gain HT | Median RMSE gain HT | Mean RMSE gain Hájek | Median RMSE gain Hájek |
|---|---|---|---|---|---|---|
| 0.7 | GBM | 6 | 0.331 | 0.291 | 0.276 | 0.346 |
| 0.6 | GBM | 6 | 0.449 | 0.474 | 0.179 | 0.256 |
| 0.4 | GBM | 6 | 0.513 | 0.477 | 0.067 | -0.088 |
| 0.7 | RF | 6 | 0.909 | 0.935 | -0.31 | -0.293 |
| 0.6 | RF | 6 | 0.695 | 0.793 | -0.204 | 0.02 |
| 0.4 | RF | 6 | 0.574 | 0.518 | -0.218 | -0.434 |

*Notes: RMSE gain is defined as RMSE without cross fitting minus RMSE with cross fitting. Positive values therefore indicate an improvement from cross fitting.*

Table 5 shows that the gains from cross fitting vary substantially across data generating scenarios, estimators, and propensity estimate methods. For the HT, cross fitting consistently improves RMSE across all scenarios and methods, with particularly pronounced gains for RF. In contrast, the benefits for the Hájek estimator are more heterogeneous and depend strongly on the underlying method. While GBM generally yields positive RMSE gains for Hájek, RF shows limited or negative gains across the considered scenarios. A similar pattern emerges for absolute bias, with cross fitting providing more consistent improvements for HT, whereas its effect on the Hájek estimator is less uniform. Overall, these results indicate that the benefits of cross fitting are strongly method and estimator dependent and are also influenced by the underlying data generating mechanism.

*Table 5. Cross fitting gains across data generating scenarios and machine learning methods.*

| Data Generating scenario | Model | Number of configurations | Mean RMSE gain HT | Mean RMSE gain Hájek | Mean absolute bias gain HT | Mean absolute bias gain Hájek |
|---|---|---|---|---|---|---|
| interaction | GBM | 6 | 0.523 | 0.25 | 0.515 | 0.075 |
| interaction | RF | 6 | 0.734 | -0.305 | -0.108 | -0.626 |
| linear | GBM | 6 | 0.302 | 0.088 | 0.284 | -0.315 |
| linear | RF | 6 | 0.72 | -0.08 | -0.525 | -0.301 |
| nonlinear | GBM | 6 | 0.468 | 0.183 | 0.456 | -0.245 |
| nonlinear | RF | 6 | 0.724 | -0.346 | 0.063 | -0.582 |

*Notes: Positive RMSE gains indicate lower RMSE under cross fitting. Positive absolute bias gains indicate lower absolute bias under cross fitting.*

Table 6 reports the best performing method according to the Hájek RMSE across sample sizes, data generating scenarios, and response rates. For the smaller sample size, GBM with five fold cross fitting is the preferred method in most configurations, particularly under linear and nonlinear data generating scenarios, while logistic regression without cross fitting is selected in some interaction settings. As the sample size increases, logistic regression without cross fitting becomes the preferred method across all data generating scenarios and response rates. Notably, RF is not selected as the best performing method in any of the considered configurations, despite its competitive performance in some settings. This pattern suggests that the relative advantage of flexible machine learning methods and cross fitting is more pronounced in smaller samples, whereas the simpler parametric specification performs better as the sample size increases. Overall, the results highlight that the optimal propensity estimation strategy depends on both the sample size and the underlying data generating mechanism.

*Table 6. Best performing method according to Hájek RMSE.*

| Sample size | Data generating scenario | Response rate | Model | Cross-fitting | Bias | Absolute bias | Variance | RMSE |
|---|---|---|---|---|---|---|---|---|
| 1000 | interaction | 0.7 | LOGIT | none | -0.003 | 0.003 | 0.24 | 0.489 |
| 1000 | interaction | 0.6 | GBM | cf5 | -0.177 | 0.177 | 0 | 0.177 |
| 1000 | interaction | 0.4 | LOGIT | none | -0.242 | 0.242 | 0.498 | 0.746 |
| 1000 | linear | 0.7 | GBM | cf5 | -0.194 | 0.194 | 0 | 0.194 |
| 1000 | linear | 0.6 | GBM | cf5 | -0.499 | 0.499 | 0 | 0.499 |
| 1000 | linear | 0.4 | GBM | cf5 | -0.315 | 0.315 | 0 | 0.315 |
| 1000 | nonlinear | 0.7 | GBM | cf5 | -0.172 | 0.172 | 0 | 0.172 |

| Sample size | Data generating scenario | Response rate | Model | Cross-fitting | Bias | Absolute bias | Variance | RMSE |
|---|---|---|---|---|---|---|---|---|
| 1000 | nonlinear | 0.6 | GBM | cf5 | -0.345 | 0.345 | 0.01 | 0.36 |
| 1000 | nonlinear | 0.4 | GBM | cf5 | -0.177 | 0.177 | 0 | 0.177 |
| 5000 | interaction | 0.7 | GBM | cf5 | -0.16 | 0.16 | 0.002 | 0.165 |
| 5000 | interaction | 0.6 | LOGIT | none | -0.039 | 0.039 | 0.056 | 0.239 |
| 5000 | interaction | 0.4 | LOGIT | none | -0.225 | 0.225 | 0.099 | 0.387 |
| 5000 | linear | 0.7 | LOGIT | none | -0.002 | 0.002 | 0.048 | 0.22 |
| 5000 | linear | 0.6 | LOGIT | none | 0.004 | 0.004 | 0.048 | 0.22 |
| 5000 | linear | 0.4 | LOGIT | none | 0.015 | 0.015 | 0.075 | 0.274 |
| 5000 | nonlinear | 0.7 | LOGIT | none | -0.001 | 0.001 | 0.047 | 0.217 |
| 5000 | nonlinear | 0.6 | LOGIT | none | -0.01 | 0.01 | 0.055 | 0.234 |
| 5000 | nonlinear | 0.4 | LOGIT | none | -0.028 | 0.028 | 0.072 | 0.27 |

*Notes: For each sample size, data generating scenario, and response rate, the method with the lowest Hájek RMSE is reported. CF denotes five fold cross fitting.*

Taken together, the simulation results point to three main findings. First, the Hájek estimator is considerably more stable than the HT estimator, particularly when flexible models such as RF are used. Second, five fold cross fitting improves RMSE in the large majority of configurations and its gains tend to be larger under lower response rates and more complex learning settings. Third, although the choice of models varies across simulation scenarios, five fold cross fitting is a remarkably consistent component of the best performing procedures.

## Conclusion

Overall, the simulation results indicate that the performance of nonresponse adjusted estimators depends substantially on the combination of the propensity score estimation method, cross fitting, response rate, and data generating scenario. Across the considered configurations, the Hájek estimator generally exhibits lower RMSE and absolute bias than the Horvitz Thompson estimator, indicating greater stability and robustness to variability in the estimated nonresponse weights.

The results also highlight important differences across machine learning methods. GBM generally provides the most favorable performance when the response propensity is estimated using cross fitting, particularly under interaction and nonlinear response mechanisms. Logistic regression performs competitively, especially in settings where the propensity model is relatively well aligned with the underlying response mechanism. In contrast, RF tends to produce substantially more variable weights, as reflected by its lower effective sample size, and this is accompanied by poorer estimation performance in several configurations.

Cross fitting does not uniformly improve performance across all methods and estimators. Its benefits are most evident for GBM, where it can reduce RMSE and absolute bias, whereas for RF the gains are less consistent and may even be negative for the Hájek estimator. This suggests that the value of cross fitting is method dependent rather than universally beneficial.

Finally, the analysis of the best performing configurations shows that the optimal method varies according to sample size, response rate, and data generating scenario. GBM with five fold cross fitting is frequently selected in the smaller sample settings and under more complex response mechanisms, whereas logistic regression without cross fitting tends to perform best in the larger sample settings. Taken together, these findings suggest that cross fitting should not be regarded as an automatic improvement, but rather as a methodological choice whose effectiveness depends on the learning algorithm and the characteristics of the response mechanism. The results further support the use of the Hájek estimator as a generally more stable alternative when estimated nonresponse propensities generate substantial weight variability.